# Ethical Hacking for IoT Security:
# A First Look into Bug Bounty Programs and Responsible Disclosure

Aaron Yi Ding
TU Delft
aaron.ding@tudelft.nl

Gianluca Limon De Jesus
TU Delft
gianluca.limon@hotmail.it

Marijn Janssen
TU Delft
m.f.w.h.a.janssen@tudelft.nl

## ABSTRACT

The security of the Internet of Things (IoT) has attracted much attention due to the growing number of IoT-oriented security incidents. IoT hardware and software security vulnerabilities are exploited affecting many companies and persons. Since the causes of vulnerabilities go beyond pure technical measures, there is a pressing demand nowadays to demystify IoT "security complex" and develop practical guidelines for both companies, consumers, and regulators. In this paper, we present an initial study targeting an unexplored sphere in IoT by illuminating the potential of crowdsource ethical hacking approaches for enhancing IoT vulnerability management. We focus on Bug Bounty Programs (BBP) and Responsible Disclosure (RD), which stimulate hackers to report vulnerability in exchange for monetary rewards. We carried out a qualitative investigation supported by literature survey and expert interviews to explore how BBP and RD can facilitate the practice of identifying, classifying, prioritizing, remediating, and mitigating IoT vulnerabilities in an effective and cost-efficient manner. Besides deriving tangible guidelines for IoT stakeholders, our study also sheds light on a systematic integration path to combine BBP and RD with existing security practices (e.g., penetration test) to further boost overall IoT security.

## 1 Introduction

Being an interface between physical and digital world, the Internet of Things (IoT) is revolutionizing our daily living by bringing interconnected services and automation to our proximate surroundings. There has been tremendous growth recently in IoT ranging from smart plugs to smart refrigerators. On one hand, this has made the old "joke" of an Internet-connected toaster become a looming reality. One the other hand, with such new omnipresence of Internet connected infrastructure, its security, safety and reliability are turning into a crucial concern for our digitalized society. For example, consider cars that have turned into computers on wheels, fitness apps that provide more health advice than our doctors, and fridges to order milk before all storage run out. As applications of IoT are becoming so diverse, ranging from eHealth, smart homes, smart industries, to smart city infrastructure, securing these pervasive services is deemed as a major societal challenge. This urge is aggregated in recent years given the growing number of IoT-oriented security incidents in terms of vulnerability exploits and data leakage [1, 2, 3, 4].

Despite the advance of security practices in IT software industry, the diversity and complexity of IoT ecosystem have made it a daunting task especially for companies to tackle IoT specific vulnerabilities in their products. Besides existing software/network security methods to minimize the risks of insecure IoT products, we need more *agile vulnerability management*, i.e., to identify the vulnerabilities as swiftly as possible to prevent attacks to occur. In this regard, as part of cyber security practices, a few organizations have started to implement Bug Bounty Programs (BBP) and Responsible Disclosure (RD) for vulnerability management purpose. Being part of the ethical hacking approaches, BBP and RD are crowdsource security practices that involve hackers to report vulnerabilities to companies in exchange for monetary rewards or credit. BBP are often initiated by companies, whereas RD are programs for sharing security breach found by ethical hackers. As revealed by Berkeley researchers, such rewarding programs do entail good potential of being economically efficient and potent when comparing to hiring full-time security specialists [5].

Research on the adoption of such crowdsource security methods for IoT is still scarce. In particular, the state of adoption, benefits, limitations, and barriers for adopting BBP and RD in IoT deserve a thorough investigation. Hence, we dive into this yet unexplored sphere where there are deficient guidelines nor insights regarding how to apply crowdsource approaches of BBP and RD for improving IoT vulnerability management.

As being the first qualitative study on the potential of using BBP and RD to enhance IoT vulnerability management, our main contributions are as follows:

- We identify causes for the lack of security practices in IoT from socio-technical and commercial perspectives. By revealing the hidden pitfalls and blind spots, stakeholders can avoid repeating the same mistakes.
- We derive a set of recommendations as best-practices that can benefit IoT vendors, developers and regulators. Considering the value chain and overall landscape in IoT, the guidelines are handy in particular for business to consumer (B2C) domain by covering how to integrate traditional mechanisms together with BBP and responsible disclosure.

We note that this study does not intend to cover all the potential vulnerabilities in IoT nor to provide an absolute solution to cure the IoT security puzzle. Instead, our focus is to provide practical and tangible recommendations and potentially new options for stakeholders to tackle IoT-oriented vulnerabilities in consumer goods, which will help enhance the overall IoT security practices. The rest of paper is organized as follows. Section 2 describes the research motivation and background. Section 3 illustrates the organization of our qualitative study on BBP and RD. Section 4 presents our recommendation and further discussions. Section 5 concludes the work.

## 2 Motivation and Background

### 2.1 Research Motivation

In this work we aim to tackle this main research quest, **"How can we enhance IoT vulnerability management practice with Bug Bounty Programs and Responsible Disclosure?"**. Our focus is on stakeholders in particular the companies and vendors in the consumer goods sector, which are in pressing demand for viable and cost-efficient IoT security solutions.

To present the overall landscape and depict the core challenge, we scrutinize the subject through three sub research questions:

1. What are the reasons for the lack of security practices by IoT developers, manufacturers, and vendors in the consumer goods sector?
2. What are the potential best practices to enhance vulnerability management for IoT products?
3. How can we leverage BBP and RD and integrate them with conventional security solutions to boost IoT security?

Given the research gap where the adoption of crowdsource security methods in IoT is still left open, our work shall contribute to the literature of security practices for IoT.

### 2.2 IoT Security Challenges

The IoT is an infrastructure of interconnected objects, people, systems and information resources together with intelligent services to allow them to process information of the physical and the virtual world and react on the inputs [6]. By collecting and exchanging data to facilitate autonomous decisions and actions, IoT is playing an increasingly prominent role in daily life. Besides the convenience, we have to face the fact that nowadays IoT is not a separate domain but is becoming part of every domain. This poses security problems and digital threats to economic growth, safety and freedom.

Given that security solutions might not keep up with all the new technological developments of IoT applications, such growing gap has made IoT security a tough challenge. There are a number of academic works covering the main problem areas for IoT [1-5]. Because of pervasive deployment, IoT applications can obtain sensitive data from individuals and businesses, and this data is a lucrative target for hackers. Under these conditions, attacks on IoT vulnerabilities can result in major safety risks. According to recent studies, IoT are at higher risks of exposure for six major reasons from the system and device perspective [7-10]:

- IoT systems do not have well-defined perimeters as they can continuously change.
- IoT systems are highly heterogeneous with respect to communication medium and protocols.
- IoT systems often include devices that are not designed to be connected to the Internet or for being secure.
- IoT devices can often autonomously control other IoT devices without human supervision.
- IoT devices could be physically unprotected and/or controlled by different parties.
- The large number of devices increases the security complexity.

Practical challenges for IoT also include that enterprises targeting end-users do not have security as a priority and are generally driven by time-to-market instead of by security requirements. Moreover, IoT security might be beyond the understanding of average IT leader's skill sets. Another important aspect is that several IoT products are the results of an increasing number of startup companies that have entered this market recently. This vast majority of startups accounts for less than 10 employees, and their obvious priority is to develop functional rather than secure products [11]. In this context, investing in security can be perceived as a costly and time-consuming obstacle. In addition, consumers demand for security is low, and they tend to prefer cheaper rather than secure products. As a result, companies lack explicit incentives to invest in security. The combination of these factors has unfortunately led to the weak security profile for many IoT systems and products, and what we often witness is the constantly newly discovered vulnerabilities that can render networks openly exploitable by breaking into a single component [7-11].

### 2.3 Vulnerability Management and Ethical Hacking

In the area of cyber security, *vulnerability management* is a pro-active security approach. The objective is to ensure vulnerabilities are identified and fixed across the product's life cycle. As show in Figure 1, the procedures of vulnerability management include the checking, identifying, verifying, mitigating and patching.

At the moment, the most conventional method, as used widely in IT industry, consists of penetration testing (Pen Testing). *Pen Testing* is a particular form of ethical hacking, which involves different phases including the identification of entry points, attempting to break in, and reporting back the discoveries [12]. Traditional Pen Tests can be performed by certified Pen Testing firms, independent ethical hackers and consultants, or even hackers hired by companies. Since the Pen Testing services are commonly paid per hour regardless of the result, penetration tests demand a great deal of money out of the company's budgets.

| Vulnerability Management Process | |
|---|---|
| Checking For Vulnerabilities | This process should include regular network scanning, firewall logging, penetration testing or use of an automated tool like a vulnerability scanner. |
| Identifying Vulnerabilities | This involves analyzing network scans and pen test results, to find anomalies that suggest a malware attack or other malicious event have taken advantage of a security vulnerability, or could possibly do so. |
| Verifying Vulnerabilities | This process includes evaluating whether the identified vulnerabilities could actually be exploited on servers. This also includes classifying the severity of a vulnerability and the level of risk. |
| Mitigating Vulnerabilities | This is the process of figuring out how to prevent vulnerabilities from being exploited before a patch is available, or in the event that there is no patch. |
| Patching Vulnerabilities | This is the process of getting patches and applying them to all the affected areas. |

**Figure 1: Vulnerability Management Process**

In recent years, Ethical Hacking has gained popularity in vulnerability management. In security context, the term "hacking" refers to unauthorized intrusion into a computer or a network, and typically has a negative connotation. However, Ethical Hacking envisages cybersecurity experts, referred as white hats or security researchers, who attack a computer system on behalf of its owners, in order to identify vulnerabilities that malicious hackers could exploit. *Ethical hackers* operate without malicious intent, but to report vulnerabilities to improve the security measures of organizations [13].

Given the fast rise of cyber-crimes over all types of commercial and government organizations, Ethical Hacking have been found powerful, and are gradually recommended by EC Council, to fight against security threats, which is in line with the statement from EC Council, "Government agencies and business organizations today are in constant need of ethical hackers to combat the growing threat to IT security. A lot of government agencies, professionals and corporations now understand that if you want to protect a system, you cannot do it by just locking your doors" [14]. The benefits of Ethical Hacking are highlighted as follows:

- Preventing data from being stolen and misused by malicious attackers.
- Discovering vulnerabilities from an attacker's point of view so to fix weak points.
- Implementing a secure network that prevents security breaches.
- Protecting networks with real-world assessments.
- Gaining the trust form customers and investors by ensuring security.
- Defending national security by protecting data from terrorism.

In terms of methodology, companies can adopt the assistance of Ethical Hacking in two ways, as shown in Figure 2. The first type is Pen Test, and the second type is crowdsource security methods such as Bug Bounty Programs and Responsible Disclosure.

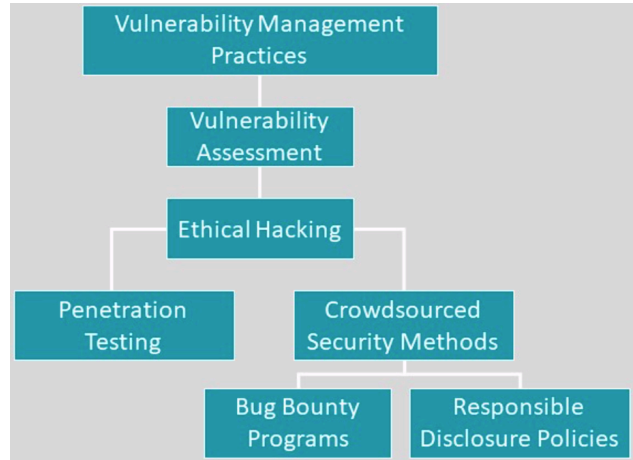

**Figure 2: Ethical Hacking in Vulnerability Management**

### 2.4 Bug Bounty Programs and Responsible Disclosure

As illustrated in Figure 2, crowdsourced security methods are the alternative for Pen Testing in Ethical Hacking. The crowdsource methods involve the participation of large numbers of ethical hackers, reporting vulnerabilities to companies in exchange for rewards that can consist of money or just recognition.

In software industry, similar practices have been utilized at large scale. One major example are the *Vulnerability Rewards Programs* (VRP) which have been applied to Chrome and Firefox, yielding several lessons on software security development. As end results, the Chrome VRP has cost approximately $580,000 over 3 years and has resulted in 501 bounties paid for the identification of security vulnerabilities. The Firefox VRP has cost approximately $570,000 over the last 3 years and has yielded 190 bounties [5].

As previous studies suggest, crowdsource methods are valid practices to identify vulnerabilities and they usually resulted in cost-effective outcome for the organizations [15, 16]. For such merits, the Gartner report indicates that approximately 50% of companies will employ by 2022 [17].

| Penetration Testing | | Crowdsourced Testing |
|---|---|---|
| Days-Weeks. Variable due to delays in finding available pen testers and appropriate skills | Onboarding | Testing community can be active within 1 hour |
| Point-In-Time. | Time Frame | Extreme flexibility. From week-long to continuous year-long testing window |
| 1-2 pen testers with variable skillsets | Personnel | Hundreds of ethical hackers from around the globe |
| Pen testers paid per hour regardless of results | Payment Model | Ethical hackers paid only for valid vulnerabilities |
| 10-60 man hours of active testing in 1 week period | Testing Hours | Hundreds of hours of active testing in 1 week period |
| One cumulative report hand-off | Vulnerability Reporting | Reporting in real-time and highly prioritized final report |

**Figure 3: Comparison of Pen Test and Crowdsource Methods**

To illustrate the difference between two types of Ethical Hacking, we highlight the details of both Pen Testing and crowdsourcing approaches in Figure 3. From a process view, Pen Testing consists of one or two people conducting security tests for a limited period of time, at the expense of a fixed amount of payment from a company. Meanwhile, crowdsource methods embrace potentially thousands of hackers working on a security target. From the scale perspective, crowdsource methods represent a more effective and efficient way to reduce security risk. In specific, instead of a point-in-time test, crowdsourcing enables continuous testing. In addition, as compared with Pen Testing, crowdsourcing hackers are only paid when a valid vulnerability is reported.

**Bug Bounty Programs (BBP)**: also known as Bug Bounties, BBP represent a reward driven crowdsource security testing where ethical hackers who successfully discover and report the vulnerabilities to companies are rewarded. The objective of BBPs is to prevent vulnerabilities from being discovered and exploited by malicious hackers and to enhance corporate security. Nowadays there are two different implementations for bug bounties: public and private programs. Public programs essential allow entire communities of ethical hackers to participate in the program. They typically consist of large scale bug bounty programs and can be both time-limited and open-end. The private programs, on the other hand, are generally limited to a selected sub-group of hackers, scoped to specific targets, and limited in time. These programs usually take place through commercial bug bounty platforms, where hackers are selected based on reputation, skills, and experience. A practical implementation of BBP often entails five steps, including assessment, preparation, internal appointment, launch, and post bounty mitigation. Currently, the main platform vendors include HackerOne[1], BugCrowd[2], Cobalt Labs[3], and Synack[4]. Those platforms have facilitated the establishing and maintaining BBPs for organizations.

**Responsible Disclosure (RD)**: also known as coordinated vulnerability disclosure, RD consists of rules and guidelines from companies that allow individuals to report vulnerabilities to organizations. The RD policies will define the models for a controlled and responsible disclosure of information upon vulnerabilities discovered by users. In this respect, most of the software vulnerabilities are typically discovered by both benign users and ethical hackers [18]. In many situations, individuals might feel responsible for reporting the vulnerability to the organization, but companies may lack a channel for them to report the found vulnerabilities. Therefore, three different outcomes might occur, including failed disclosure, full disclosure, and organization capture. Among these three, the target of RD is the organization capture where companies create a safe channel and provide rules for submitting vulnerabilities to their security team, and further allocate resources to follow up the process. Figure 4 depicts the work flow of RD in practice.

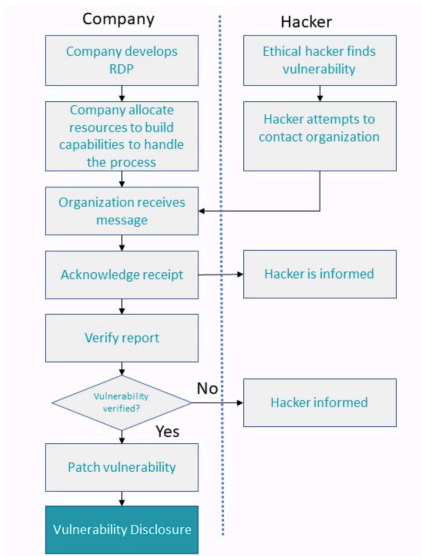

**Figure 4: Work Flow of Responsible Disclosure**

Being the major two types of crowdsource methods, the main difference between BBP and RD is that BBP involves an actual invitation from companies to hackers to hack their systems in exchange for monetary rewards, whereas RD does not invite hackers to hack but are instead coordinating the disclosure of security breaches that ethical hackers identify. Moreover, RD generally does not offer monetary rewards to hackers.

## 2.5 Knowledge Gap of Ethical Hacking for IoT

In conventional IT industry, Ethical Hacking has been used for software and network security. In particular, BBP and RD have broadened the range of crowdsource Ethical Hacking solutions. However, based on our extensive literature review and interview (detailed in Section 3), there are insufficient studies on how to properly conduct Ethical Hacking for IoT.

The first reason for such gap is IoT hardware. As IoT systems are highly dependent on specific hardware to realize dedicated functions, the hardware vulnerability is a critical spot. However, once an IoT product is delivered to the market, it is very challenging for companies to fix vulnerabilities in the hardware. Normally, after a vulnerability is identified, companies are able to release software updates to improve the security of a device. In the case of IoT, it is often not possible to do the same to fix vulnerabilities in the hardware in a scalable and low-cost way.

The second reason is lack of incentive to invest in security. This reason partially comes from the attitude of IoT consumers, which indirectly affect the companies' motivation to adopt efficient mechanisms such as BBP and RD to offer more secure products.

---

[1] https://www.hackerone.com/
[2] https://www.bugcrowd.com/
[3] https://cobalt.io/
[4] https://www.synack.com/

A further problem is that most IoT devices are produced with cheap components from low-cost suppliers. There is a severe lack of awareness about the importance of testing such hardware, as vendors often fail to understand that hardware is an important attack vector.

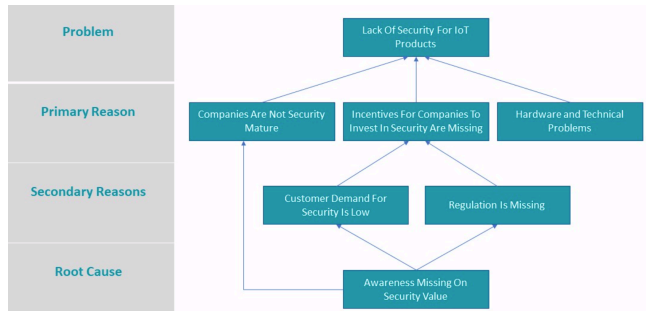

Figure 5: Reasons for Lack of IoT Security Practice

We highlight in Figure 5 the reasons for the lack of security practices in IoT. These reasons reflect a cyclical problem of digital technologies where the initial phase is on improving functionality and expanding its applications. In such phase, security is considered as a cost. This insight serves as a primary hint for introducing cost-efficient crowdsource methods such as BBP and RD to IoT, since the features of BBP and RD match well with the hidden demands behind the identified reasons for lack of IoT security.

## 3 Qualitative Study and Analysis

### 3.1 Survey and Expert Interviews

Given the exploratory nature of this study, our research consists of extensive literature survey (conveyed in Section 2) and semi-structured interviews with academic and industrial experts.

In our semi-structured interviews, we prepared 14 questions that cover the IoT security risks, perceptions and measures, and also the application of crowdsource security methods in IoT. We repeated the same structure across the interviews, and when additional themes become relevant to be asked, more open discussions are applied. The majority of the interviews were realized as face-to-face with the experts, and in the rest of the cases done by remote video interviews. All of the interviews were recorded with the consent of experts, in order to preserve the integrity of the data. Each interview lasted for approximately 30 to 45 minutes.

Regarding expert selection, we utilized the judgment sampling, where only a limited number of subjects in a population have the chance to be selected, based on their expertise in the topic to be investigated. In our case of IoT security, we define an "expert" as an individual with considerable expertise in the field of cybersecurity, and also possess deep knowledge on IoT security hacking, including ethical hacking practices and/or crowdsourced methods.

We were able to consult with 19 experts, coming from 9 different companies, spread across 5 different European countries. The experts hold different years of experience in different industries and have different roles and levels within the company.

### 3.2 Analysis of Results

To systematically process the interview data, our approach consists of three main steps: data reduction, data display and the drawing of conclusions, following the qualitative analysis guidelines [19].

For data reduction, all the audio recordings of the interviews were transcribed and converted into written documents. Subsequently, the next step consisted of coding the data. Our coding entails iterative process of labeling words, sentences or entire paragraphs to reduce and rearrange the data in a meaningful way. Once the codes are generated from all the transcripts, we conclude the data reduction process with categorization. As a result, the codes were organized and categorized in different groups. For our research, the coding and categorization were realized adopting ATLAS.ti, a computer program for the qualitative analysis of large bodies of textual data.

As part of the data reduction, 2 out of the 19 were excluded from the data analysis. The reason is that two of the companies, namely Automated IoT Security Analyses Platform and Multinational Electronics Company, did not have the required expertise in IoT security and ethical hacking as the other ones in our study. In the end, 17 transcripts were used in the data analyses.

For data display, we extrapolated categories and created conceptual groups from the coding, and hence determine the frequency in the data for each category counting the number of experts that were referring to the same conclusion.

On drawing conclusions from the qualitative data to provide answers to the research questions, we based our statement on the identification of common themes in the data reduction. By linking the categories, comparing them, and seeking contrasts with the data in the literature, we attempt to provide a logical explanation for the observed patterns and to determine our conclusions (in Section 4).

## 4 Recommendations and Discussions

### 4.1 Best Practices for Integrating BBP and Responsible Disclosure

To tackle our research question "what are the potential best practices to enhance vulnerability management for IoT products?", our main recommendation is that companies should implement security from the design of products. This practice makes easier the process of vulnerability management because once those companies realize security from the beginning, it is less likely that they will face vulnerabilities. This is highlighted by one of our interviewees as "Security should be a parallel process. You build in some tests in your software development life cycle, you do code reviews, you do secure design, and you have security requirements. Then whenever there is a release you do pen testing to make sure you follow a methodical process that validates that the security requirements are indeed implemented. But then in parallel with all of it, you also have a responsible disclosure policy on your website together with a bug bounty. This way you'll be able to cover other classes of vulnerabilities that pen testers missed."

From an organization perspective, one recommendation is to allocate the security budget from IT to the risk department of organizations, since keeping the budget for security within the IT is nowadays insufficient to cover the IoT vulnerabilities and the consequences across the entire organization.

In specific to BBP, incentives for both ethical hackers and companies are the most crucial element. In this respect, companies shall take initiative to provide more incentives to researchers through the means of sharing hardware and setting up credit/reputation system to recognize active participation. In implementation, BBP shall be applied after Pen Test to achieve a balance between cost and efficiency of detecting vulnerabilities.

For RD, we recommend the implementation of RD policies to all type of companies, in particular for those dealing with IoT. For companies, being open-minded is a condition to utilize crowdsource ethical hacking. The next step is to define policies for what happens after a vulnerability has being reported. In addition, companies need to define the internal processes to be able to accept and process the incoming reports.

### 4.2 Adoption Strategy for IoT Companies

To tackle our research question, "how can we leverage BBP and RD and integrate them cost-efficiently with conventional security solutions to boost IoT security?", we derived a set of guidelines that cover both technical and organizational aspects.

For IoT hardware challenge, given that shipping IoT hardware such as a smart car to testers is generally not feasible, we recommend organizing hacking events which can take the form of a live one to two-day hacking events. Such IoT white hat hacking event can connect the team from company together with the "crowdsourced" white-hat hackers in a highly interactive environment to accelerate the discovery of critical vulnerabilities in IoT hardware and software. One recent example of such an event was the car hacking event organized by BugCrowd in Louisville of Kentucky[5].

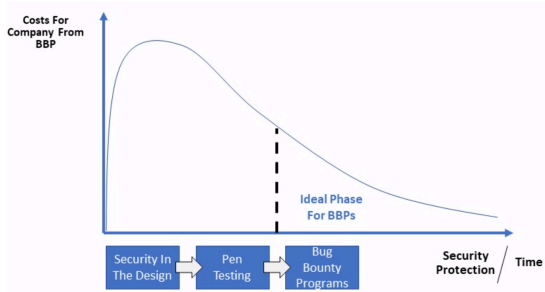

**Figure 6: Cost and Adoption Phase for BBP**

---
[5] https://www.bugcrowd.com/blog/the-first-ever-car-hacking-event-of-its-kind-bcbugbash-louisville/

Offering private BBP is another strategy besides public platforms. In particular for companies that do not want to have the burden to manage the whole process themselves, they can operate their BBPs using intermediary platforms. To illustrate the cost and stage of adoption, Figure 6 presents our suggestion of adoption sequence of security design, integrating conventional methods (Pen Test), with ethical hacking approaches. BBP is used as an example in this figure. Based on Figure 6, our recommendation is to launch BBP after a certain state of maturity. Before starting with BBP, companies should already be active with vulnerability management practices to ensure that a sufficient amount of vulnerabilities have been previously identified and fixed. Otherwise, BBP can turn into a very expensive and ineffective approach for companies to identify vulnerabilities. The reason for this is that companies might have developed a product and never tested it. Untested products typically have many vulnerability issues. When BBP is used immediately on this product, hackers will start reporting a large amount of vulnerabilities and for each vulnerability there is a price. It is possible that a company will pay a very high amount in a very short period of time, much more than what to be paid for a Pen Test.

### 4.3 Limitations

In this qualitative study, one limitation is that only experts that were conveniently available participated in the interview. Within judgmental sampling, there is a risk of selection bias. Therefore, the generalizability of our findings to the entire population should be done with care. In addition, it is worth noting that our sample is based on a population of security experts working as security advisors. Consequently, our research takes the point of view of experts that are generally providing solutions to companies that face security problems. We did not yet include in our study experts from other types of companies, and they might have a different view on the problems.

Due to the fact that the experts that participated in the research were predominantly from the Netherlands (13 experts), and more in general all from Europe, the results should be generalized with care to other countries and the whole security industry. In fact, security practices are different among countries. Some countries are advanced whereas others are still learning. In the western world, cybersecurity is generally mature but there are still many regions where cybersecurity is not mature and responsible disclosure and bug bounties are just too sophisticated for them at the moment.

### 5 Conclusion

This study provides a first look into the potential of BBP and RD for IoT vulnerability management. Besides sharing insights on integration path, we also reveal hidden pitfalls and blind spots that deserve special attentions from IoT stakeholders. As a solid step to demystify the potential of crowdsource ethic hacking for IoT, our work tackled the major research question, "**How can we enhance the IoT vulnerability management practice with Bug Bounty**

**Programs and Responsible Disclosure?**". For IoT vulnerability management, our recommendation is to launch BBP only after companies have performed initial security testing and fixed the problems. The objective of BBP and RD policies should always be to provide additional support in finding undetected vulnerabilities, and never to be the only security practice. Follow-up procedures also need to be defined when vulnerabilities are found. Our findings raise the awareness on IoT security and presents a call for further actions from companies, consumers and regulators in the IoT B2C domain.

## ACKNOWLEDGMENTS

We would like thank Prof. Mark Alfano for his suggestions on interview analysis in this research.

## REFERENCES

[1] DigiCert. 2018. State of IoT Security Survey 2018. Retrieved from https://www.digicert.com/wpcontent/uploads/2018/11/StateOfIoTSecurity_Report_11_02_18_F_am.pdf.
[2] Saleh Soltan, Prateek Mittal, H. Vincent Poor. 2018. BlackIoT: IoT Botnet of High Wattage Devices Can Disrupt the Power Grid. In Proceedings of the 27th USENIX Security Symposium (USENIX Security 18).
[3] J. Granjal, E. Monteiro and J. Sá Silva, "Security for the Internet of Things: A Survey of Existing Protocols and Open Research Issues," in IEEE Communications Surveys & Tutorials, vol. 17, no. 3, pp. 1294-1312, 2015.
[4] Cisco Visual Networking Index: Forecast and Trends, 2017–2022 White Paper.
[5] Matthew Finifter, Devdatta Akhawe, and David Wagner. 2013. An empirical study of vulnerability rewards programs. In Proceedings of the 22nd USENIX conference on Security (SEC '13).
[6] ISO/IEC 20924:2018 Information technology -- Internet of Things (IoT) -- Vocabulary.
[7] I. Hafeez, A. Y. Ding, L. Suomalainen, A. Kirichenko, S. Tarkoma. Securebox: Toward Safer and Smarter IoT Networks. In Proceedings of ACM CoNEXT Workshop on Cloud-Assisted Networking (CAN '16).
[8] I. Hafeez, A. Y. Ding, S. Tarkoma. 2017. IOTURVA: Securing Device-to-Device (D2D) Communication in IoT Networks. In Proceedings of the 12th ACM MobiCom Workshop on Challenged Networks (CHANTS '17).
[9] Ibbad Hafeez, Aaron Yi Ding, Markku Antikainen, Sasu Tarkoma. 2018. Real-Time IoT Device Activity Detection in Edge Networks. In: Au M. et al. (eds) Network and System Security. NSS 2018. Lecture Notes in Computer Science, vol 11058. Springer, Cham.
[10] Bertino, E., & Islam, N., (2017). Botnets and internet of things security. Computer, (2), 76-79.
[11] N. Zhang, et al., (2017). Understanding IoT security through the data crystal ball: Where we are now and where we are going to be. arXiv preprint arXiv:1703.09809.
[12] Baloch, R., (2014). Ethical hacking and penetration testing guide. Auerbach Publications.
[13] V. Chandrika, (2014). Ethical hacking: Types of ethical hackers. International Journal of Emerging Technology in Computer Science & Electronics (IJETCSE), 11(1).
[14] EC-Council, (2019). What is Ethical Hacking | Types of Ethical Hacking | EC-Council. Retrieved from https://www.eccouncil.org/ethical-hacking/#
[15] Laszka, A., Zhao, M., & Grossklags, J. (2016). Banishing Misaligned Incentives for Validating Reports in Bug-Bounty Platforms. Computer Security – ESORICS 2016 Lecture Notes in Computer Science,161-178. doi:10.1007/978-3-319-45741-3_9.
[16] Laszka, A., Zhao, M., Malbari, A., & Grossklags, J. (2018). The rules of engagement for bug bounty programs. Extended version http://aronlaszka. com/papers/laszka2018rules. pdf (February 2018).
[17] Gartner, Inc., (2018). Emerging Technology Analysis: Bug Bounties and Crowdsourced Security Testing.
[18] H. Cavusoglu, S. Raghunathan, (2005). Emerging Issues in Responsible Vulnerability Disclosure. In Workshop on Information Technology and Systems.
[19] Miles, M. B., & Huberman, A. M., (1994). Qualitative data analysis an expanded sourcebook. Thousand Oaks: SAGE.